# Importing Relationships into a Running Graph Database Using Parallel Processing

Joshua R. Porter, Aleks Y. M. Ontman

**Abstract** - Importing relationships into a running graph database using multiple threads running concurrently is a difficult task, as multiple threads cannot write information to the same node at the same time. Here we present an algorithm in which relationships are sorted into bins, then imported such that no two threads ever access the same node concurrently. When this algorithm was implemented as a procedure to run on the Neo4j graph database, it reduced the time to import relationships by up to 69% when 32 threads were used.

## 1 Introduction

Graph databases have been rising in prominence in recent years because of their natural suitability for storing data sets in which information about connections between entities matters at least as much as that about the entities themselves [1]. A graph database is organized as nodes connected by relationships. In such cases, when the organization of the database matches that of the data, a graph database allows a user to write simpler queries that execute faster than if the data were stored in a traditional relational database.

One common challenge with graph databases is the need for rapid data import, reducing the "time to graph." It is often possible to import nodes into a running graph database using parallel processing, provided that the import can be organized such that no two threads are ever writing information to the same node at the same time, i.e., the isolation property of database transactions is satisfied [2]. When importing a list of data about nodes in which no node is ever mentioned more than once, this condition is trivially satisfied. If a list of data about nodes contains more than one reference to the same node, the list can be sorted by the reference so that multiple references to the same node are handled by the same thread, avoiding a conflict.

Importing relationships into a running graph database using parallel processing is much more difficult. For an arbitrary list of data about relationships between nodes, there is no simple way to sort the data by node reference to avoid conflicts between threads, as each relationship has references to two nodes. Consequently, a list of data about relationships between nodes is generally imported using a single thread, which avoids conflicts but does not fully utilize the power of modern multicore hardware.

The algorithm described here is designed to circumvent this problem. A list of data about relationships is sorted into a two-dimensional array of bins based on the identifiers of the source and target nodes. The relationships in these bins are imported in a series of rounds; in each round, multiple threads import relationships from a set of bins such that no two threads are ever accessing the same nodes. We implemented this algorithm as a procedure to run on the Neo4j graph database and found that it significantly reduced the time to import relationships into the database, consistent with expectations.

## 2 Algorithm description

The goal is to import relationships into a graph database using 2n threads running concurrently. The following description will illustrate this proposed method for the case of n = 2 (4 threads).

Binning relationships

The first step is to create a two-dimensional set of bins, 2n+1 by 2n+1. For the example of n = 2, this amounts to an 8 x 8 grid, shown at right.

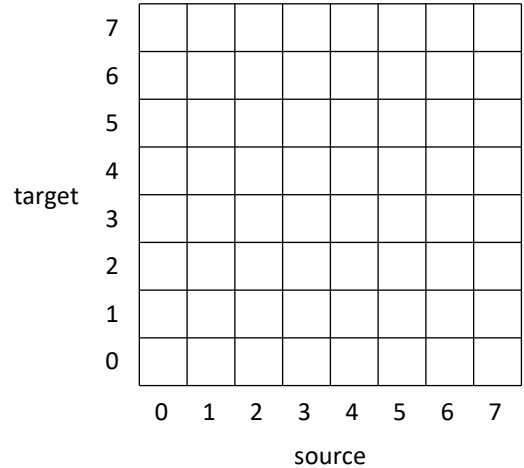

When a relationship is created between a source node and a target node, each node is first located using an identifier, which can be an integer (e.g., an ID number), a string (e.g., a title), or some other data type. The source and target node identifiers are used to determine the bin for the relationship as follows:

- If the identifier is not an integer, compute an integer hash code for the identifier, then take the rightmost n + 1 bits of that hash code.
- If the identifier is an integer, simply take its rightmost n + 1 bits. This is a preferable situation, as a hash code is more computationally expensive.

For the case of n = 2, the rightmost 3 bits of the identifier (or its hash code) for the source node are used; this corresponds to an integer between 0 and 7, inclusive. Similar steps are followed for the target node, and thus we determine which of the bins in the grid above should receive this relationship.

## 3 Orchestrating the import of bins

The relationships are now imported by bin in a series of rounds. In each round, a single thread imports the relationships in a single bin. Thus, since there are 2n threads and 2$^{2n+2}$ bins, there must be 2n+2 rounds. In each round, no thread may access any node that could possibly be accessed by any other thread running in the same round, as this would result in an error. For example, if in a given round thread #1 is importing relationships from bin (3,5), no other thread may import relationships from any bin with x- or y-coordinate 3 or 5, as the relationships in any such bin might have a source node or target node being accessed concurrently by thread #1.

The following pseudocode generates a list of bins to import in each round such that the above rule is satisfied. It first determines which bins are imported in each pair of rounds, then divides the bins in each roundPair between two individual rounds roundPair.A and roundPair.B.

- For roundPair = 0 to 2n+1 – 1:
  - For x = 0 to 2n+1 – 1:
    - Let y = x XOR roundPair (the bitwise XOR operation is equivalent to flipping the bits of x specified by roundPair)
    - If y > x, then add the bin (x,y) to round roundPair.A
    - If y < x, then add the bin (x,y) to round roundPair.B
    - If y = x, then add the bin (x,y) to either round roundPair.A or roundPair.B, wherever there is room

For the example of n = 2, the following table shows the roundPairs and the bins (x,y) generated by the XOR operation for each roundPair.

| roundPair | Bins (x, y = x XOR roundPair) |
| --- | --- |
| 0 (binary 000) | (0,0), (1,1), (2,2), (3,3), (4,4), (5,5), (6,6), (7,7) |
| 1 (binary 001) | (0,1), (1,0), (2,3), (3,2), (4,5), (5,4), (6,7), (7,6) |
| 2 (binary 010) | (0,2), (1,3), (2,0), (3,1), (4,6), (5,7), (6,4), (7,5) |
| 3 (binary 011) | (0,3), (1,2), (2,1), (3,0), (4,7), (5,6), (6,5), (7,4) |
| 4 (binary 100) | (0,4), (1,5), (2,6), (3,7), (4,0), (5,1), (6,2), (7,3) |
| 5 (binary 101) | (0,5), (1,4), (2,7), (3,6), (4,1), (5,0), (6,3), (7,2) |
| 6 (binary 110) | (0,6), (1,7), (2,4), (3,5), (4,2), (5,3), (6,0), (7,1) |
| 7 (binary 111) | (0,7), (1,6), (2,5), (3,4), (4,3), (5,2), (6,1), (7,0) |

When the bins (x,y) are divided into individual rounds roundPair.A and roundPair.B as specified in the pseudocode, the result is as follows:

| roundPair | Round | Bins |
|---|---|---|
| 0 | 0.A | (0,0), (1,1), (2,2), (3,3) |
| | 0.B | (4,4), (5,5), (6,6), (7,7) |
| 1 | 1.A | (0,1), (2,3), (4,5), (6,7) |
| | 1.B | (1,0), (3,2), (5,4), (7,6) |
| 2 | 2.A | (0,2), (1,3), (4,6), (5,7) |
| | 2.B | (2,0), (3,1), (6,4), (7,5) |
| 3 | 3.A | (0,3), (1,2), (4,7), (5,6) |
| | 3.B | (2,1), (3,0), (6,5), (7,4) |
| 4 | 4.A | (0,4), (1,5), (2,6), (3,7) |
| | 4.B | (4,0), (5,1), (6,2), (7,3) |
| 5 | 5.A | (0,5), (1,4), (2,7), (3,6) |
| | 5.B | (4,1), (5,0), (6,3), (7,2) |
| 6 | 6.A | (0,6), (1,7), (2,4), (3,5) |
| | 6.B | (4,2), (5,3), (6,0), (7,1) |
| 7 | 7.A | (0,7), (1,6), (2,5), (3,4) |
| | 7.B | (4,3), (5,2), (6,1), (7,0) |

This result has two important properties:

1. Each coordinate in the grid is listed exactly once.
2. In each round, any given coordinate 0-7 is present in no more than one bin (x,y).

Thus, when using 2n threads, relationships can be imported in a series of 2n+2 rounds. In each round, one thread imports the relationships that sort into one bin. Because of property #2, no two threads will ever try to access the same node at the same time. Because of property #1, all relationships will be imported.

It is important to note that to achieve optimum performance the relationships must be approximately uniformly distributed between the bins. In some cases, the sorting process may be modified (e.g., by changing how the hash code is calculated) to make the distribution of relationships more uniform. In other cases, such as a collection of relationships that all connect to the same node, it may be impossible to distribute the relationships between bins with any uniformity, regardless of the sorting function. In the worst case, the performance reduces to that of a single-threaded relationship import, with a small penalty for the computational effort required to sort relationships into bins.

## 4 Results

This algorithm was implemented in Java as a procedure to run on the Neo4j graph database. The implementation, called iterateRelationship(), was based on the apoc.periodic.iterate() procedure for Neo4j; the original code was modified to implement binning and relationship import by round as described above. The procedure was tested using an implementation of Neo4j Enterprise Edition 3.5.4 running on an Amazon Web Services EC2 instance (r5a.8xlarge, 32 virtual CPUs, 256 GB RAM, 25 GB elastic block storage) with Ubuntu Linux 16.04.5. Two random graphs were generated to test data import performance: an Erdős-Rényi random graph (independent, equally likely relationships) with 5,000,000 nodes and 7,500,000 relationships, and a Barabási-Albert random graph (relationships preferentially attached) with 5,000,000 nodes and 9,999,996 relationships [3, 4]. For each graph, a CSV file of nodes and another CSV file of relationships were generated. After the nodes were imported into a new graph, the relationships were imported using either apoc.periodic.iterate() with a single thread or iterateRelationship() with 2n threads for several values of n. A range of batch sizes was tested; each combination of n and batch size was tested three times, and the time taken to import the relationships was recorded.

Under certain conditions, the iterateRelationship() procedure generated errors related to deadlocks on nodes, despite the algorithm's design to prevent deadlocks. This may be due to a known issue with the Neo4j lock manager1; the suggested workaround is to retry any transaction that fails. Thus, iterateRelationship() includes functionality for a user-defined number of retries, which allowed it to run successfully.

Figure 1 shows the time taken to import the relationships in the Erdős-Rényi random graph into the database for different numbers of threads and batch sizes; Figure 2 shows the same for the Barabási-Albert random graph. In general, using more threads decreased the time to import the relationships, though the return on additional threads diminished as the number of threads increased. Moreover, for each number of threads, there was an optimal batch size yielding a minimal import time; this optimal batch size decreased as the number of threads increased. For the Erdős-Rényi graph, the minimal import time with 32 threads was 69% less than that with a single thread; for the Barabási-Albert graph, it was 67% less.

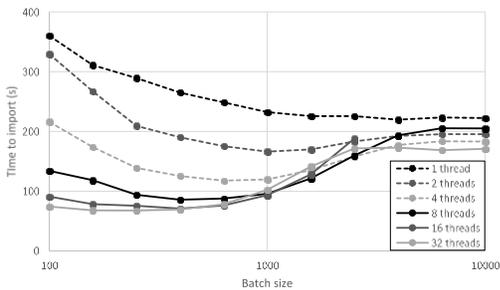

Figure 1: Time to import 7,500,000 relationships from an Erdős-Rényi random graph into the database using different numbers of threads and batch sizes. Single-threaded cases used apoc.periodic.iterate(); multithreaded cases used iterateRelationship(). All measurements represent the average of n=3 trials.

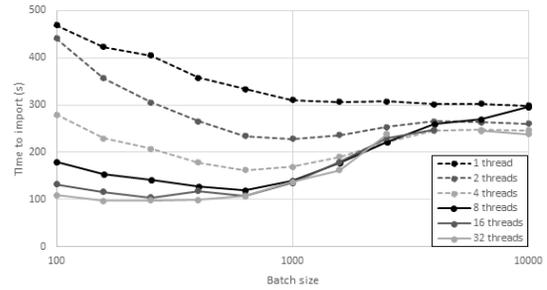

Figure 2: Time to import 9,999,996 relationships from a Barabási-Albert random graph into the database using different numbers of threads and batch sizes. Single-threaded cases used apoc.periodic.iterate(); multithreaded cases used iterateRelationship(). All measurements represent the average of n=3 trials.

## 5 Discussion

The algorithm described here was able to significantly reduce the time to import a large number of relationships into a running graph database. By dividing relationships into bins and importing bins of relationships in a series of rounds such that no two threads can access the same nodes at the same time, this method enables a graph database to take full advantage of the power of modern multicore hardware when importing relationships while satisfying the requirements of transactionality.

The implementation of this algorithm as a procedure for Neo4j could benefit greatly from certain improvements. It is likely that the deadlock errors and the consequent transaction retries slowed relationship import considerably over what is theoretically possible. Changes to the lock manager could address this issue and enable even larger speed improvements, especially if it becomes possible to import larger batch sizes without errors.

This algorithm could potentially be applied in other graph database contexts as well. As graph databases grow in popularity, this algorithm could enable them to process larger volumes of data at faster rates, enhancing their ability to provide real-time business intelligence.

---

[1] https://github.com/neo4j/neo4j/issues/12040

**Joshua R. Porter** received his B.S. in electrical and computer engineering from Lafayette College in 2006 and his Ph.D. in electrical and computer engineering from Johns Hopkins University in 2012. From 2012-2017, he did postdoctoral work in the National Cancer Institute, researching the p53 tumor suppressor protein and using mathematical models to understand its function. He is currently a Senior Consultant with Deloitte Consulting LLP, using graph databases to organize data and yield valuable insights.

**Aleks Y. M. Ontman** received his B.S. in electrical engineering from SUNY Binghamton in 2005 and his Ph.D. in Materials Science and Engineering from the University of Virginia in 2012. He is currently a manager in Strategy & Analytics practice with Deloitte Consulting LLP, specializing in application of Natural Language Processing and graph theory to Government and commercial clients.